\documentclass[11pt]{article}
\usepackage{hyperref}
\pdfoutput=1
\usepackage[utf8x]{inputenc}
\usepackage{graphicx}

\title{Coherent Structures in Turbulent Flow over Two-Dimensional River Dunes}
\author{Mohammad Omidyeganeh and Ugo Piomelli \\
\\\vspace{6pt} Department of Mechanical and Materials Engineering, \\ Queen's University, Kingston, Canada}

\begin{document}

\maketitle

We performed large-eddy simulations of the flow over a typical
two-dimensional dune geometry at laboratory scale (the Reynolds number
based on the average channel height and mean velocity is 18,900) using
the Lagrangian dynamic eddy-viscosity subgrid-scale model \cite{OmidyeganehP11b,OmidyeganehP11a}.  The
governing differential continuity and Navier-Stokes equations are
discretized on a nonstaggered grid using a second order in time and
space curvilinear finite-volume code.

\begin{figure}[h!]
  \centering
  \includegraphics[width=0.8\linewidth]{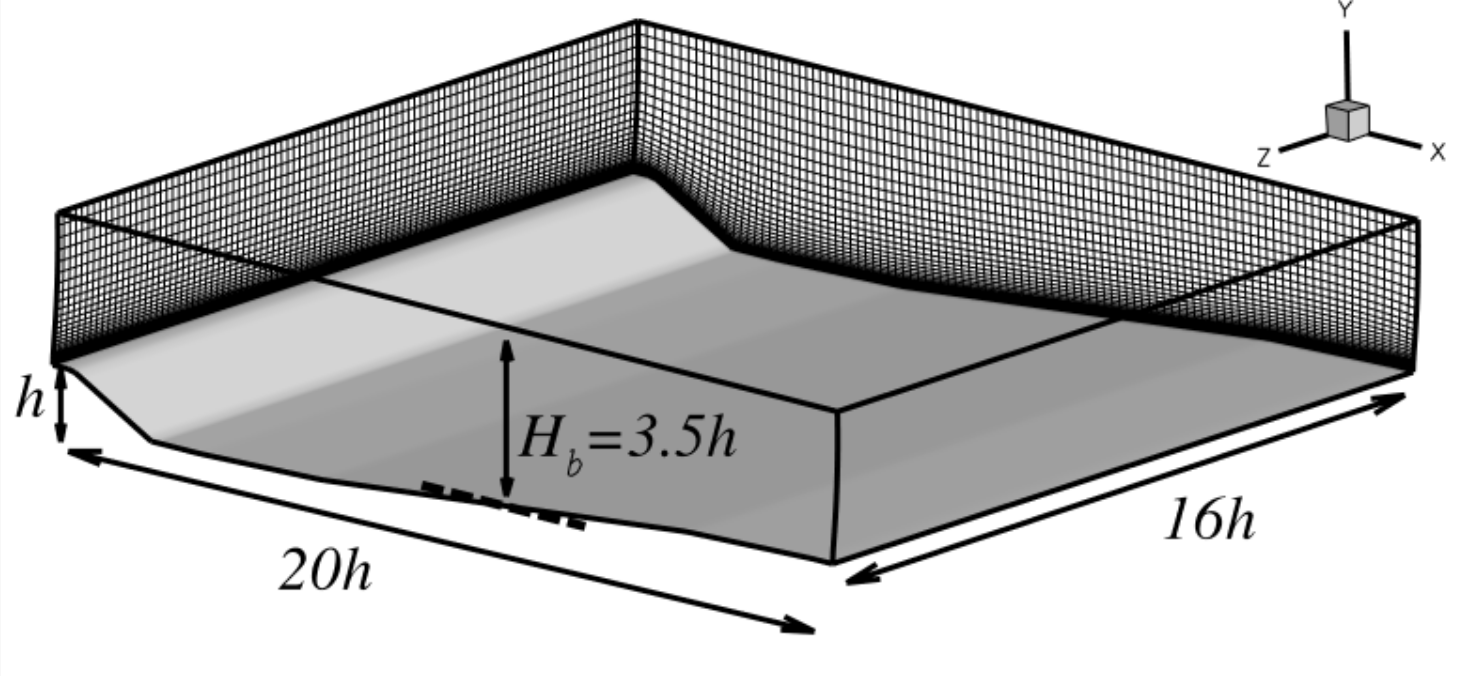}
  % contour.png: 849x253 pixel, 72dpi, 29.95x8.93 cm, bb=0 0 849 253
  \caption{Sketch of the physical configuration. Every fourth grid line is shown.}
  \label{fig:geometry}
\end{figure}
The computational configuration is sketched in Figure
\ref{fig:geometry}.  Periodic boundary conditions are used in the
streamwise ($x$) and spanwise ($z$) directions.  At the free surface,
the wall normal velocity is set to zero, as are the vertical
derivatives of the streamwise and spanwise velocity components.  An
orthogonal mesh with $416\times128\times384$ grid points is used,
which results in grid spacings in local wall units $\Delta s^{+}
\approx 12.9$ (streamwise), $\Delta z^{+}\approx 6.0$ (spanwise), and
$0.1<\Delta n^{+} < 12.1$ (wall normal).

The flow separates at the dune crest and reattaches downstream on the
bed (at $x\simeq 5.7h$) \cite{OmidyeganehP11a}. A favorable pressure gradient accelerates
the flow over the stoss-side (the upward-sloping region for $x >8h$)
and an unfavorable gradient for $x<8h$ decelerates the flow over the
lee-side of the dune. Due to the separation of the flow, a shear layer
is generated after the crest that expands in the wake region towards
the next dune.

In the \href{https://qshare.queensu.ca/Users01/8mo9/www/MO-UP-DFD64-Low.mpg}{video}, the outer-layer
turbulence structures are visualized through isosurfaces of pressure
fluctuations colored by distance to the surface. Spanwise vortices
are generated in the shear layer separating from the crest due to the
Kelvin–Helmholtz instability. They are convected downstream and either
interact with the wall or rise to the surface, taking the form of
large horseshoe-like structures. These structures may undergo an
intense distortion, become one-legged, or be completely destroyed.  As
they grow to dimensions comparable to (or larger than) the flow depth,
a strong ejection occurs between their legs. The interaction between
two large coherent structures may result in merging, or in dissipation
of the weaker one. Surviving eddies grow and rise along the shear
layer emanating from the dune crest; they tilt downward and,
eventually, their tips touch the surface. When the legs of the
horseshoe are close to the surface, they create an upwelling, which
expands and weakens. The legs of the vortex loop remain coherent for
a longer time \cite{OmidyeganehP11b,OmidyeganehP11a}.

The video showing the full description of the setup of the problem and
the results can be seen at the following URLs:
\begin{itemize}
 \item \href{https://qshare.queensu.ca/Users01/8mo9/www/MO-UP-DFD64-Low.mpg}{Low resolution} 
 \item \href{https://qshare.queensu.ca/Users01/8mo9/www/MO-UP-DFD64-High.mpg}{High resolution} 
\end{itemize}

This video has been submitted to the Gallery of Fluid Motion 2011
which is an annual showcase of fluid dynamics videos. More detailed analysis
on the characteristics of turbulence structures are included in references \cite{OmidyeganehP11b,OmidyeganehP11a}.

The authors thank the Natural Sciences and Engineering Research
Council (NSERC) under the Discovery Grant program and the High
Performance Computing Virtual Laboratory (HPCVL), Queen’s University
site, for the computational support.  Mohammad Omidyeganeh
acknowledges the partial support of NSERC under the Alexander Graham
Bell Canada NSERC Scholarship Program.  Ugo Piomelli also acknowledges
the support of the Canada Research Chairs Program.

\bibliographystyle{plain}
\bibliography{ref}

\end{document}